\pdfoutput=1
\documentclass[twoside]{zHenriquesLab-StyleBioRxiv}

\usepackage{graphicx}
\usepackage{booktabs}
\usepackage{amsmath}
\usepackage[numbers,sort&compress]{natbib} 

\leadauthor{Qi}
\begingroup
\begin{document}

\title{Normal and Atypical Mitosis Image Classifier using Efficient Vision Transformer}
\shorttitle{AMF/NMF mitosis classification}

\author[1]{Xuan Qi}
\author[2]{Dominic Labella}
\author[3]{Thomas Sanford}
\author[1]{Maxwell Lee}

\affil[1]{Laboratory of Cancer Biology and Genetics, NCI, NIH, Bethesda, MD, 20852, USA}
\affil[2]{Department of Radiation Oncology, Duke University Medical Center, Durham, NC, 27705, USA}
\affil[3]{University of Hawaii Cancer Center, Honolulu, HI, 96813, USA}

\maketitle

\begin{abstract}
We tackle atypical versus normal mitosis classification in the MIDOG 2025 challenge using EfficientViT-L2, a hybrid CNN--ViT architecture optimized for accuracy and efficiency. A unified dataset of 13,938 nuclei from seven cancer types (MIDOG++ and AMi-Br) was used, with atypical mitoses comprising $\sim$15\%. To assess domain generalization, we applied leave-one-cancer-type-out cross-validation with 5-fold ensembles, using stain-deconvolution for image augmentation. For challenge submissions, we trained an ensemble with the same 5-fold split but on all cancer types. In the preliminary evaluation phase, this model achieved balanced accuracy of 0.859, ROC AUC of 0.942, and raw accuracy of 0.85, demonstrating competitive and well-balanced performance across metrics.
\end{abstract}


\begin{corrauthor}
qix3@nih.gov
\end{corrauthor}

\section{Background}
\label{sec:background}
\subsection{datasets}
We combined two officially provided datasets: Midog and Ami-Br into a single unified dataset and removed all duplicated samples to ensure data integrity. The final dataset comprises 13,938 annotated histopathology image samples spanning seven cancer types—four from canine tumors and three from human cancers, as summarized in Table~\ref{tab:dataset}. Each sample is labeled at the nucleus level as either Atypical Mitosis Figure (AMF) or Normal Mitosis Figure (NMF). Across the entire dataset, there exists very obvious class imbalance: AMF samples constitute only 8\% to 24\% of the nuclei within each cancer type, with an overall average of 15.6\%. This imbalance poses a significant challenge for training robust classifiers and motivates the need for effective handling strategies such as data augmentation or class-weighted loss functions.

\begin{table}[h!]
\centering
\resizebox{\linewidth}{!}{%
\begin{tabular}{lcccc}
\toprule
\textbf{Cancer Type} & \textbf{Sample Num} & \textbf{AMF} & \textbf{NMF} & \textbf{AMF \%} \\
\midrule
Canine cutaneous mast cell tumor & 2327 & 351 & 1976 & 15.10\% \\
Canine lung cancer & 855 & 110 & 745 & 12.90\% \\
Canine lymphoma & 3959 & 317 & 3642 & 8.00\% \\
Canine soft tissue sarcoma & 1286 & 210 & 1076 & 16.30\% \\
Human breast cancer & 3722 & 832 & 2890 & 22.40\% \\
Human melanoma & 1150 & 271 & 879 & 23.60\% \\
Human neuroendocrine tumor & 639 & 85 & 554 & 13.30\% \\
\midrule
\textbf{Total} & \textbf{13938} & \textbf{2176} & \textbf{11762} & \textbf{15.60\%} \\
\bottomrule
\end{tabular}%
}
\caption{Distribution of AMF and NMF samples across different cancer types.}
\label{tab:dataset}
\end{table}

\section{Methods}

\subsection{Efficient Vision Transformer (ViT)}

Vision Transformers (ViTs) are well suited for image classification as their self-attention mechanism captures long-range dependencies and global context from the very first layer, enabling integration of both fine-grained local features and broader structural cues. For mitosis classification, model efficiency is equally critical since whole-slide images contain millions of patches, requiring fast inference and low memory use to ensure practical deployment in clinical workflows. To address both accuracy and efficiency, we adopt EfficientViT \cite{liu2023efficientvit}, which combines convolutional and transformer modules in a hybrid architecture and employs cascaded linear attention to reduce computation and memory cost. Specifically, we use EfficientViT-L2 with a 256×256 input size, achieving 85.37$\%$ ImageNet accuracy with 64M parameters and 9.1 GMACs, offering an effective balance between performance and efficiency for large-scale mitosis classification.

\subsection{Leave one out cross-validation (LOOCV) and model Ensemble}

In real-world diagnostic settings, models often encounter unseen data that may differ significantly from the training data distribution. This challenge is especially relevant in our case, as the dataset used in this study includes only 7 cancer types, far fewer than the full spectrum of known cancers. To evaluate the generalization of our approach under such a practical and challenging scenario, we adopt a leave-one-out cross-validation (LOOCV) strategy. Specifically, we exclude one cancer type during training and use the remaining six for model development, testing the model's performance on the held-out type.

Furthermore, for each LOOCV setting where the model is trained on data from 6 cancer types, we apply 5-fold cross-validation. In each fold, 4 subsets are used for training and 1 for validation, resulting in 5 trained models per LOOCV configuration. This yields a total of 7$\times$5=35 models across all cancer types in this work. For evaluation on the held-out cancer type, we perform model ensembling by averaging the predicted AMF probability scores (ranging from 0.0 to 1.0) across the 5 models, using the mean confidence values as the final output.

Lastly, for final submission to preliminary and final stage of challenge, we train our ensemble model with same 5-fold split policy but on all cancer types data.

\subsection{Data Augmentation and Other settings}

For data augmentation, we follow the H$\&$E stain based augmentation propose in \cite{tellez2018_HE_Stain_aug}. In this approach, image patches are first transformed into H$\&$E color space via color deconvolution, after which each channel is perturbed independently by random scaling and shifting before being projected back to RGB. These controlled perturbations mimic realistic variations in stain intensity and hue while preserving tissue morphology, thereby exposing the network to a broader distribution of staining conditions.

For other deep learning related settings, we use weighted sampling for our data sampler, focal loss as loss function and AdamW as the optimizer.

\section{Results}
We first run our test with the LOOCV and 5-fold split settings mentioned in Section \ref{sec:background}. Then, we submit our 5-fold ensemble model trained with all data to the preliminary eval phase of Midog 2025 challenge.

\subsection{Performance metrics for LOOCV tests}

\begin{figure}[htbp]
    \centering
    \includegraphics[width=0.96\linewidth]{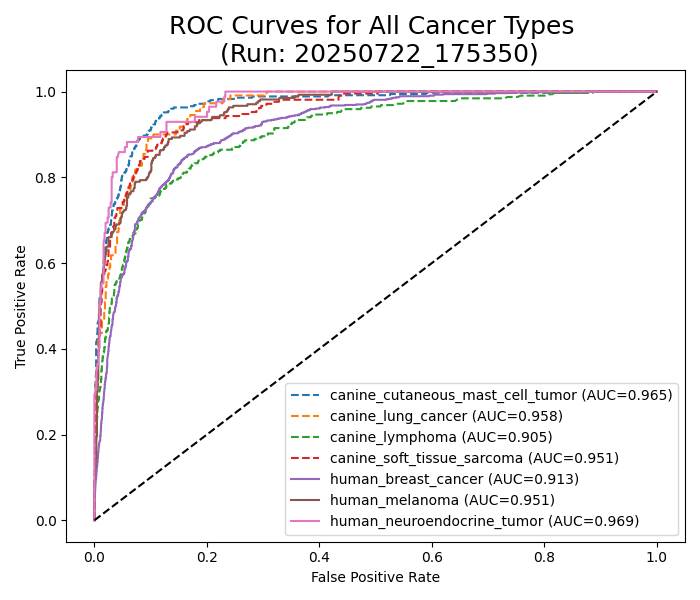}
    \caption{ROC curves for testing all cancer types}
    \label{fig:roc}
\end{figure}

We first plot the receiver operating characteristic (ROC) curves for each cancer type in Figure~\ref{fig:roc}. As a threshold-independent metric, the ROC curve reflects the model’s ability to distinguish between AMF and NMF samples. Overall, the model demonstrates strong discriminative power, with consistently high AUC values across most cancer types.

For the BA score and confusion matrix, we adopt the Otsu thresholding \cite{otsu_thres} method to automatically determine the decision threshold for each cancer type in the test set. Otsu’s method is a ground-truth–free algorithm that selects an optimal threshold by maximizing the between-class variance in the prediction scores. This approach allows us to adaptively define thresholds without relying on manual tuning or label distribution assumptions, making it suitable for evaluating performance across imbalanced data or differernt cancer types.

\begin{figure*}[htbp]
    \centering
    \includegraphics[width=\textwidth]{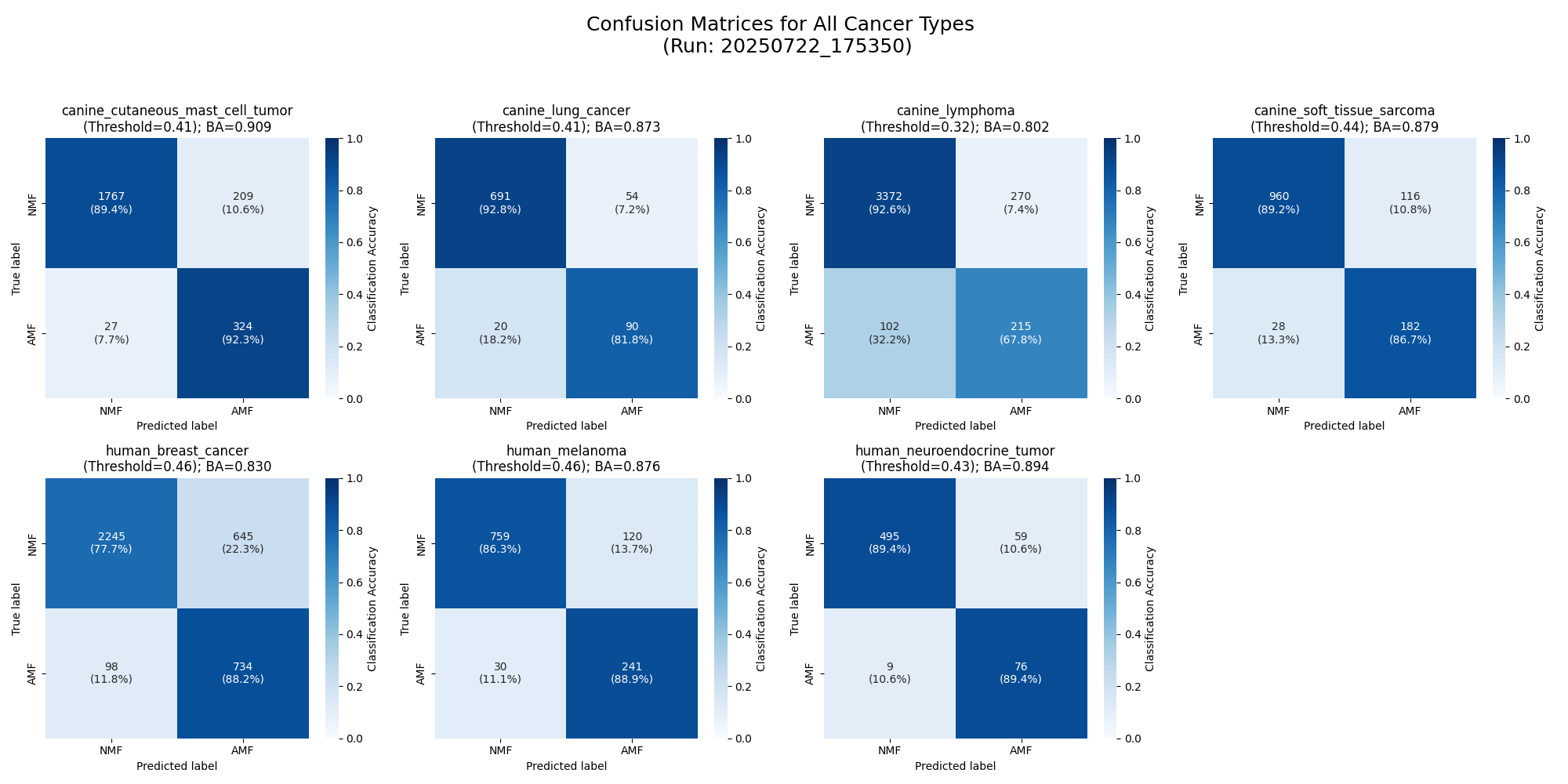}
    \caption{Confusion matrix for testing all cancer types}
    \label{fig:confusion}
\end{figure*}

Figure~\ref{fig:confusion} presents the confusion matrices for each cancer type, showing class-wise prediction distributions after applying Otsu thresholding. These matrices provide insight into model bias and class-specific accuracy. Across most cancer types, AMF and NMF samples are well-separated, with high diagonal values indicating strong correct classification.

Table~\ref{tab:cancer_type_metrics} summarizes the classification performance across cancer types using four metrics: ROC AUC, balanced accuracy (BA), NMF accuracy, and AMF accuracy. While ROC AUC reflects threshold-independent performance, the remaining metrics are calculated using thresholds automatically selected by the Otsu method.

\begin{table}[htbp]
\centering
\caption{Performance metrics by cancer type}
\resizebox{\linewidth}{!}{%
\begin{tabular}{lcccc}
\hline
\textbf{Cancer Type} & \textbf{ROC\_AUC} & \textbf{BA} & \textbf{NMF\_acc} & \textbf{AMF\_acc} \\
\hline
canine\_cutaneous\_mast\_cell\_tumor & 0.965 & 0.909 & 0.894 & 0.923 \\
canine\_lung\_cancer                  & 0.958 & 0.873 & 0.928 & 0.818 \\
canine\_lymphoma                     & 0.905 & 0.802 & 0.926 & 0.678 \\
canine\_soft\_tissue\_sarcoma        & 0.951 & 0.879 & 0.892 & 0.867 \\
human\_breast\_cancer                & 0.913 & 0.830 & 0.777 & 0.882 \\
human\_melanoma                      & 0.951 & 0.876 & 0.863 & 0.889 \\
human\_neuroendocrine\_tumor         & 0.969 & 0.894 & 0.894 & 0.894 \\
\hline
\end{tabular}%
}
\label{tab:cancer_type_metrics}
\end{table}

\subsection{Preliminary Eval Phase result}
In the preliminary evaluation phase, our model achieved consistent and well-balanced performance across all four domains. The average ROC AUC reached 0.942, demonstrating strong discriminative capability. The overall balanced accuracy was 0.859, highlighting stable performance across positive (AMF) and negative (NMF) classes. In addition, the model maintained a solid overall accuracy of 0.85, with sensitivity (0.873) and specificity (0.844) remaining well aligned.

Taken together, among high-ranked methods, our approach offers a well-balanced tradeoff across key evaluation metrics. Since the BA score is threshold dependent, it should be considered alongside threshold-independent ROC AUC, while raw accuracy provides additional context on overall performance. Under this comprehensive view, our method achieves competitive and well-balanced results across BA, ROC AUC, and raw accuracy. The raw eval results can be found \cite{midog2025_stage1_eval} and ranking info can be found \cite{midog2025_leaderboard}.

\begin{table}[htbp]
\centering
\caption{Preliminary evaluation phase results across domains}
\label{tab:prelim_eval}
\resizebox{\linewidth}{!}{%
\begin{tabular}{c|c|c|c|c|c}
\hline
\textbf{Domain} & \textbf{ROC AUC} & \textbf{Accuracy} & \textbf{Sensitivity} & \textbf{Specificity} & \textbf{BA} \\
\hline
Domain 0 & 0.875 & 0.806 & 1.000 & 0.781 & 0.891 \\
Domain 1 & 0.898 & 0.820 & 0.724 & 0.841 & 0.783 \\
Domain 2 & 0.974 & 0.888 & 0.972 & 0.854 & 0.913 \\
Domain 3 & 0.972 & 0.895 & 1.000 & 0.889 & 0.944 \\
\hline
\textbf{Overall} & \textbf{0.942} & \textbf{0.850} & \textbf{0.873} & \textbf{0.844} & \textbf{0.859} \\
\hline
\end{tabular}%
}
\end{table}

\bibliography{references}


\end{document}